\documentclass[aps,prl,twocolumn,superscriptaddress,floatfix,nofootinbib,showpacs,amsmath,amssymb,altaffilletter]{revtex4}

\usepackage{graphicx}
\usepackage{dcolumn}
\usepackage{bm}
\usepackage{color}
\usepackage{verbatim}
\usepackage{paralist}

\newcommand{\ber}{\mbox{$^{7}$Be}}
\newcommand{\bor}{\mbox{$^{8}$B}}
\newcommand{\cele}{\mbox{$^{11}$C}}
\newcommand{\cfo}{\mbox{$^{14}$C}}

\newcommand{\kr}{\mbox{$^{85}$Kr}}
\newcommand{\rb}{\mbox{$^{\rm 85}$Rb}}
\newcommand{\rbm}{\mbox{$^{\rm 85m}$Rb}}
\newcommand{\bite}{\mbox{$^{210}$Bi}}
\newcommand{\pbfo}{\mbox{$^{214}$Pb}}
\newcommand{\Bipo}{\mbox{$^{214}$Bi-$^{214}$Po}}
\newcommand{\bipo}{\mbox{$^{212}$Bi-$^{212}$Po}}
\newcommand{\tho}{\mbox{$^{232}$Th}}
\newcommand{\ura}{\mbox{$^{238}$U}}

\newcommand{\Pee}{\mbox{$P_{ee}$}}
\newcommand{\fpp}{\mbox{$f_{pp}$}}
\newcommand{\fbe}{\mbox{$f_{\rm Be}$}}
\newcommand{\fB}{\mbox{$f_{\rm B}$}}
\newcommand{\fCNO}{\mbox{$f_{\rm CNO}$}}

\newcommand{\mub}{\mbox{$\mu_B$}}

\begin{document}
\title{New results on solar neutrino fluxes from 192 days of Borexino data}

\newcommand{\APC}{Laboratoire AstroParticule et Cosmologie, 75231 Paris cedex 13, France}
\newcommand{\Budapest}{KFKI-RMKI, 1121 Budapest, Hungary}
\newcommand{\Dubna}{Joint Institute for Nuclear Research, 141980 Dubna, Russia}
\newcommand{\Genova}{Dipartimento di Fisica, Universit\`a e INFN, Genova 16146, Italy}
\newcommand{\Heidelberg}{Max-Planck-Institut f\"ur Kernphysik, 69029 Heidelberg, Germany}
\newcommand{\Kiev}{Kiev Institute for Nuclear Research, 06380 Kiev, Ukraine}
\newcommand{\Krakow}{M.~Smoluchowski Institute of Physics, Jagiellonian University, 30059 Krakow, Poland}
\newcommand{\Kurchatov}{RRC Kurchatov Institute, 123182 Moscow, Russia}
\newcommand{\LNGS}{INFN Laboratori Nazionali del Gran Sasso, SS 17 bis Km 18+910, 67010 Assergi (AQ), Italy}
\newcommand{\Milano}{Dipartimento di Fisica, Universit\`a degli Studi e INFN, 20133 Milano, Italy}
\newcommand{\MIT}{Physics Department, Massachusetts Institute of Technology, Cambridge, MA, 02139 USA}
\newcommand{\Munich}{Physik Department, Technische Universit\"at Muenchen, 85747 Garching, Germany}
\newcommand{\Pavia}{INFN, Pavia 27100, Italy}
\newcommand{\Perugia}{Dipartimento di Chimica, Universit\`a e INFN, 06123 Perugia, Italy}
\newcommand{\Peters}{St. Petersburg Nuclear Physics Institute, 188350 Gatchina, Russia}
\newcommand{\Princeton}{Physics Department, Princeton University, Princeton, NJ 08544, USA}
\newcommand{\PrincetonChemEng}{Chemical Engineering Department, Princeton University, Princeton, NJ 08544, USA}
\newcommand{\Queens}{Physics Department, Queen's University, Kingston ON K7L 3N6, Canada}
\newcommand{\UMass}{Physics Department, University of Massachusetts, Amherst, AM 01003, USA}
\newcommand{\Virginia}{Physics Department, Virginia Polytechnic Institute and State University, Blacksburg, VA 24061, USA}

\author{C.~Arpesella}\altaffiliation{deceased.}\affiliation{\LNGS}\affiliation{\Milano}
\author{H.O.~Back}\altaffiliation{Now at North Carolina State University, NC, USA.}\affiliation{\Virginia}
\author{M.~Balata}\affiliation{\LNGS}
\author{G.~Bellini}\affiliation{\Milano}
\author{J.~Benziger}\affiliation{\PrincetonChemEng}
\author{S.~Bonetti}\affiliation{\Milano}
\author{A.~Brigatti}\affiliation{\Milano}
\author{B.~Caccianiga}\affiliation{\Milano}
\author{L.~Cadonati}\affiliation{\UMass}\affiliation{\Princeton}
\author{F.~Calaprice}\affiliation{\Princeton}
% ...

\collaboration{Borexino Collaboration}
\noaffiliation

\date{\today}

\begin{abstract}
We report the direct measurement of the \ber\ solar neutrino signal rate performed with the Borexino detector at the Laboratori Nazionali del Gran Sasso.  The interaction rate of the 0.862~MeV \ber\ neutrinos is 49$\pm$3$_{\rm stat}$$\pm$4$_{\rm syst}$~counts/(day$\cdot$100~ton).  The hypothesis of no oscillation for \ber\ solar neutrinos is inconsistent with our measurement at the 4$\sigma$~C.L..  Our result is the first direct measurement of the survival probability for solar $\nu_e$ in the transition region between matter-enhanced and vacuum-driven oscillations.  The measurement improves the experimental determination of the flux of \ber, $pp$, and CNO solar $\nu_e$, and the limit on the magnetic moment of neutrinos.
\end{abstract}

\keywords{Solar neutrinos; Neutrino oscillations; Low background detectors; Liquid scintillators}
\pacs{13.35.Hb, 14.60.St, 26.65.+t, 95.55.Vj, 29.40.Mc}

\maketitle

Neutrino oscillations~\cite{bib:osc} are the established mechanism to explain the solar neutrino problem, which originated from observations in radiochemical experiments with a sub-MeV threshold~\cite{bib:rchem-cl,bib:rchem-ga} and from real time observation of high energy neutrinos~\cite{bib:kamiokande,bib:sno}.  Neutrino oscillations were also observed in atmospheric neutrinos~\cite{bib:kamiokande} and have been confirmed with observation of reactor $\bar{\nu}_e$~\cite{bib:kamland}.  Borexino is the first experiment to report a real-time observation of low energy solar neutrinos below 4.5~MeV~\cite{bib:bxfirstresults}, which were not accessible so far with the state-of-the art detector technologies because of natural radioactivity.  In this Letter we report the direct measurement of the low energy (0.862~MeV) \ber\ solar neutrinos with the Borexino detector from an analysis of 192 live days in the period from May 16, 2007 to April 12, 2008, totaling a 41.3~ton$\cdot$yr fiducial exposure to solar neutrinos.

Solar neutrinos are detected in Borexino through their elastic scattering on electrons in the scintillator.  Electron neutrinos ($\nu_e$) interact through charged and neutral currents and in the energy range of interest have a cross section $\sim$5~times larger than $\nu_\mu$ and $\nu_\tau$, which interact only via the neutral current.  The electrons scattered by neutrinos are detected by means of the scintillation light retaining the information on the energy, while information on the direction of the scattered electrons is lost.  The basic signature for the mono-energetic 0.862~MeV \ber\ neutrinos is the Compton-like edge of the recoil electrons at 665~keV.

The key features of the Borexino detector are described in Refs.~\cite{bib:bxtechnology,bib:bxfirstresults,bib:bxdetectorpaper}.  Borexino is a scintillator detector with an active mass of 278~tons of pseudocumene (PC, 1,2,4-trimethylbenzene), doped with 1.5~g/liter of PPO (2,5-diphenyloxazole, a fluorescent dye).  The scintillator is contained in a thin (125~$\mu$m) nylon vessel~\cite{bib:bxvessels} and is surrounded by two concentric PC buffers (323 and 567~tons) doped with 5.0~g/l of dimethylphthalate (DMP), a component quenching the PC scintillation light.  The two PC buffers are separated by a second thin nylon membrane to prevent diffusion of radon towards the scintillator.  The scintillator and buffers are contained in a Stainless Steel Sphere (SSS) with diameter 13.7~m.  The SSS is enclosed in a 18.0-m~diameter, 16.9-m high domed Water Tank (WT), containing 2100~tons of ultrapure water as an additional shield.  The scintillation light is detected via 2212 8'' PMTs uniformly distributed on the inner surface of the SSS~\cite{bib:bxcones,bib:bxpmts}.  Additional 208 8" PMTs instrument the WT and detect the Cherenkov light radiated by muons in the water shield, serving as a muon veto.

The key requirement in the technology of Borexino is achieving extremely low radioactive contamination, at or below the expected interaction rate of 0.5~counts/(day$\cdot$ton) expected for \ber\ neutrinos.  The design of Borexino is based on the principle of graded shielding, with the inner core scintillator at the center of a set of concentric shells of increasing radiopurity.  All components were screened and selected for low radioactivity~\cite{bib:bxrad}, and the scintillator and the buffers were purified on site at the time of filling~\cite{bib:bxpur,bib:bxlakn}.  Position reconstruction of the events, as obtained from the PMTs timing data via a time-of-flight algorithm, allows to fiducialize the active target: approximately 2/3 of the scintillator serves as an active shield.

Events are selected by means of the following cuts:

\begin{compactenum}[i]
\item Events must have a unique cluster of PMTs hits, to reject pile-up of multiple events in the same acquisition window.
\item Muons and all events within a time window of 2~ms after a muon are rejected.
\item Decays due to radon daughters occurring before the \Bipo\ delayed coincidences are vetoed.  The fraction surviving the veto is accounted for in the analysis.
\item Events must be reconstructed within a spherical fiducial volume corresponding approximately to 1/3 of the scintillator volume in order to reject external~$\gamma$~background.  Additionally, we require the $z$-coordinate of the reconstructed vertex, measured from the center of the detector, to satisfy $|z|$$<$1.7~m in order to remove background near the poles of the inner nylon vessel.
\end{compactenum}
The combined loss of fiducial exposure due to the cuts i-iii is ~0.7\%.  The fiducial cut iv results in a fiducial mass of 78.5~tons.

\begin{figure}[!t]
\includegraphics[width=0.5\textwidth]{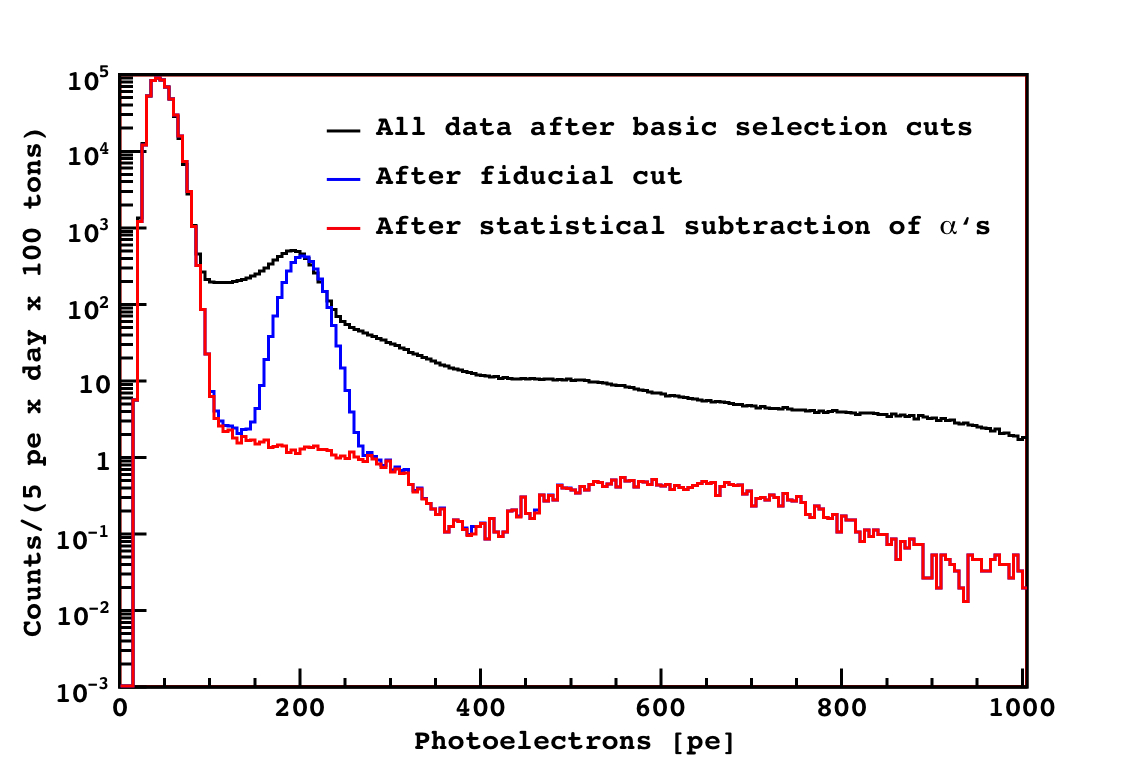}
\caption{The raw photoelectron charge spectrum after the basic cuts i--iii (black), after the fiducial cut iv (blue), and after the statistical subtraction of the $\alpha$-emitting contaminants (red).  All curves scaled to the exposure of 100~day$\cdot$ton.  Cuts described in the text.}
\label{fig:spectra}
\end{figure}

The black curve in Fig.~\ref{fig:spectra} is the spectrum of all events surviving the basic cuts i-iii: below 100~photoelectrons~(pe) the spectrum is dominated by $^{14}$C~decays ($\beta^-$, $Q$=156~keV) intrinsic to the scintillator~\cite{bib:bxc14} and the peak at 200~pe is due to $^{210}$Po~decays ($\alpha$, $Q$=5.41~MeV, light yield quenched by $\sim$13), a daughter of $^{222}$Rn out of equilibrium with the other isotopes in the sequence.  The blue curve is the spectrum after the fiducial cut iv.  The red curve is obtained by statistical subtraction of the $\alpha$-emitting contaminants, by use of the pulse shape discrimination made possible by the PC-based scintillator~\cite{bib:bxab}.  Prominent features include the Compton-like edge due to \ber\ solar neutrinos (300--350~pe) and the spectrum of \cele\ ($\beta^+$, $Q$=1.98~MeV, created in situ by cosmic ray-induced showers, 400--800~pe).

The study of fast coincidence decays of \Bipo\ (from \ura) and \bipo\ (from \tho) yields, under the assumption of secular equilibrium, contamination for \ura\ of (1.6$\pm$0.1)$\times$10$^{-17}$~g/g and for \tho\ of (6.8$\pm$1.5)$\times$10$^{-18}$~g/g.  The \kr\ content in the scintillator was probed through the rare decay sequence $\kr \to \rbm + e^+ + \nu_e$, $\rbm \to \rb + \gamma$ ($\tau$=1.5~$\mu$s, BR 0.43\%) that offers a delayed coincidence tag.  Our best estimate for the activity of \kr\ is 29$\pm$14~counts/(day$\cdot$100~ton).

\begin{figure}[!t]
\includegraphics[width=0.5\textwidth]{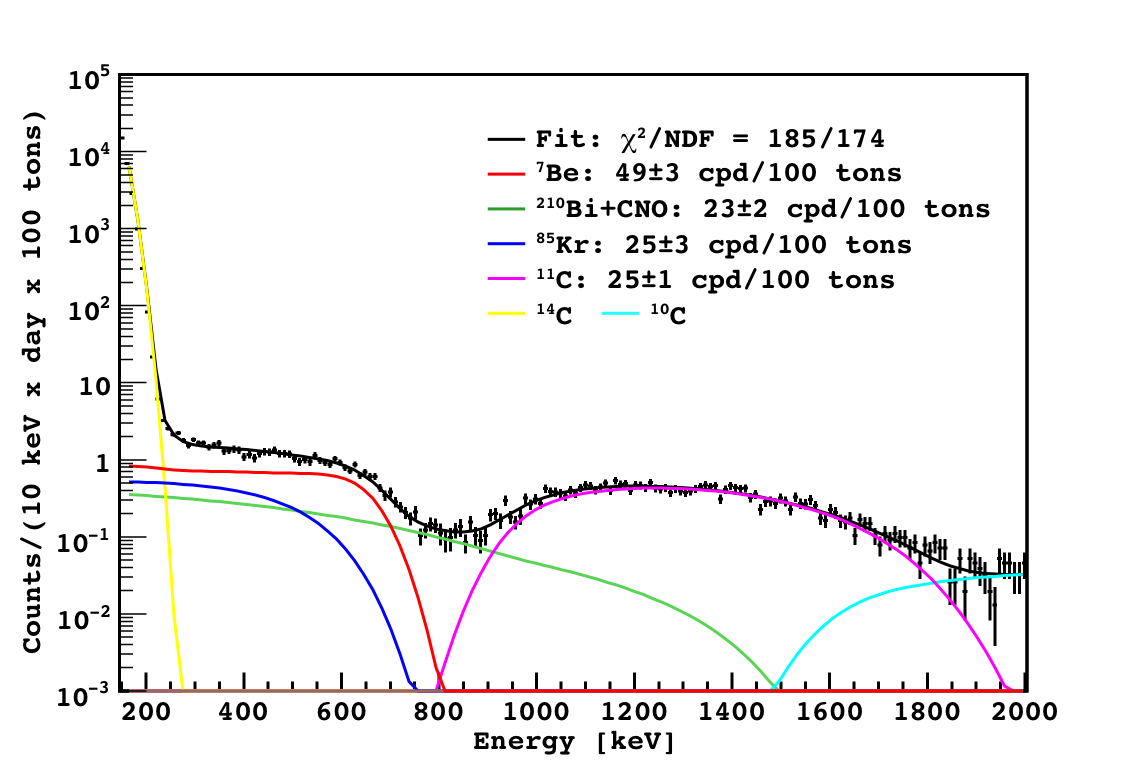}
\caption{Spectral fit in the energy region 160--2000~keV.}
\label{fig:fit}
\end{figure}

\begin{table}[!b]
\begin{center}
\caption{Fit Results [counts/(day$\cdot$100~ton)].}
\begin{tabular}{lc}
\hline\hline
\ber\			&49$\pm$3$_{\rm stat}$ \\
\kr\			&25$\pm$3$_{\rm stat}$ \\
\bite+CNO	&23$\pm$2$_{\rm stat}$ \\
\cele\		&25$\pm$1$_{\rm stat}$ \\
\hline\hline
\end{tabular}
\label{tab:fit-results}
\end{center}
\end{table}

\begin{figure}[!t]
\includegraphics[width=0.5\textwidth]{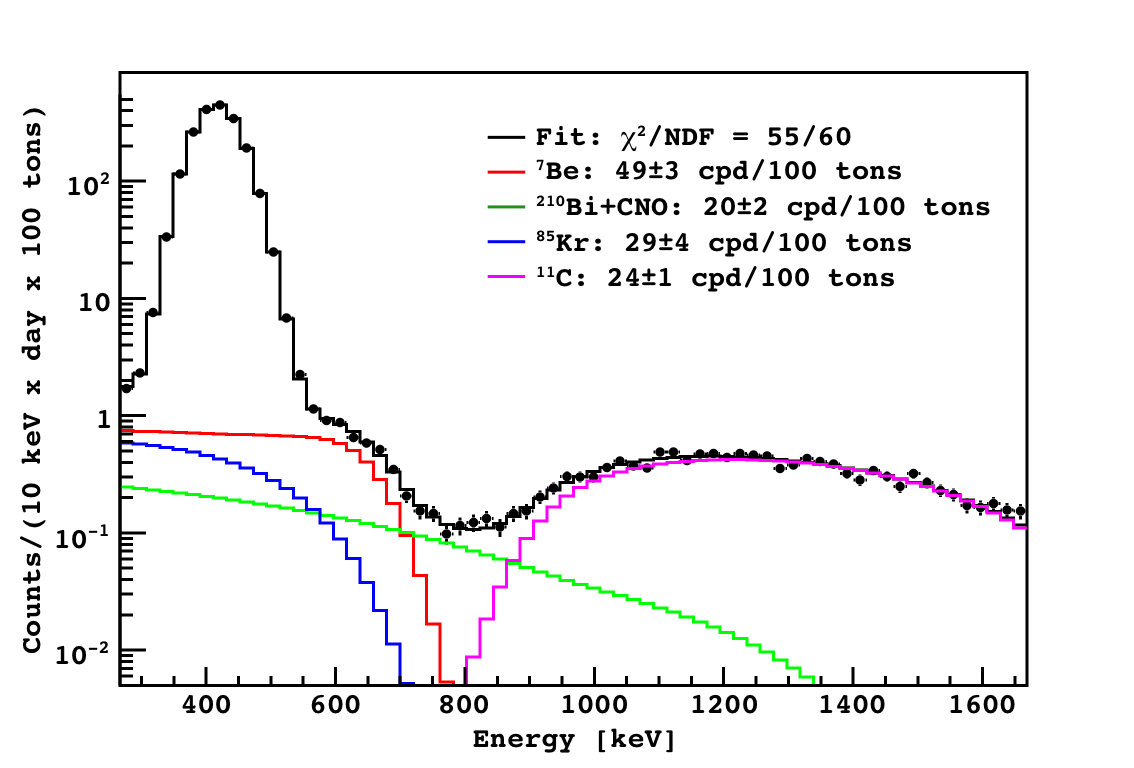}
\caption{Spectral fit in the energy region 260--1670~keV prior to statistical $\alpha$'s subtraction.}
\label{fig:fit-alpha}
\end{figure}

We determined the light yield and the interaction rate of \ber\ solar neutrinos by fitting the $\alpha$-subtracted spectrum in the region 100--800~pe, accounting for the presence of several possible contaminants.  We obtain a light yield of about 500~pe/MeV for $\beta$'s at the minimum of ionization, and the energy resolution is approximately scaling as 5\%/$\sqrt{E~{\rm [MeV]}}$.  The weights for \cfo, \cele, and \kr\ are left as free parameters in the fit.  The \pbfo\ surviving cut iii is independently determined and its weight in the fit is fixed.  Weights for {\it pp} and {\it pep} neutrinos are fixed to the values expected from the Standard Solar Model (SSM)~\cite{bib:carlos} and from a recent determination of $\sin^2{2\theta_{12}}$=0.87 and $\Delta m_{12}^2$=7.6$\times$10$^{-5}$~eV$^2$~\cite{bib:kamland}, which correspond to the Large Mixing Angle (LMA) scenario of solar neutrino oscillation via the MSW effect~\cite{bib:msw}.  The spectra for CNO neutrinos and \bite\ are almost degenerate and cannot be distinguished prior to removal of the \cele\ background~\cite{bib:hagner,bib:11c}: we use a single component whose weight is a free parameter.  Two independent analysis codes report consistent spectra and results, shown in Fig.~\ref{fig:fit} and summarized in Table~\ref{tab:fit-results}.  A further check was performed by fitting the spectrum obtained prior to statistical $\alpha$'s subtraction, obtaining consistent results, as shown in Fig.~\ref{fig:fit-alpha}.

Several sources, as summarized in Table~\ref{tab:systerr}, contribute to the systematic error.  The total mass of scintillator (315~m$^3$, 278~ton) is known within $\pm$0.2\%.  Not so yet for the fiducial mass, which is defined by a software cut.  We estimate the systematic error to be $\pm$6\% on the basis of the distribution of reconstructed vertexes of uniform background sources (\cfo, 2.2~MeV $\gamma$-rays from capture of cosmogenic neutrons, daughters of Rn introduced during the filling with scintillator) and on the basis of the inner vessel radius determined from the reconstructed position of sources located at the periphery of the active volume (\bipo\ coincidences emanating from $^{228}$Th contaminations in the nylon of the inner vessel and $\gamma$-rays from the buffer volumes).  The uncertainty in the detector response function results in a large systematic error, as small variations in the energy response affect the balance of counts attributed by the fit to \ber\ and \kr.  We aim at reducing substantially the global systematic uncertainty with the forthcoming deployment of calibration sources in the detector: this will allow a 3D mapping of the performance of position reconstruction algorithms and an in-depth study of the detector response function as a function of $\beta$- and $\gamma$-ray energies.

\begin{table}[!b]
\begin{center}
\caption{Estimated Systematic Uncertainties [\%].}
\begin{tabular}{lD{.}{.}{2.1}lD{.}{.}{3.2}}
\hline\hline
Total Scintillator Mass			&0.2		&Fiducial Mass Ratio			&6.0 \\
Live Time						&0.1		&Detector Resp. Function			&6.0	\\
Efficiency of Cuts				&0.3		& 							& \\
\hline
\multicolumn{3}{l}{Total Systematic Error}	& 8.5 \\
\hline\hline
\end{tabular}
\label{tab:systerr}
\end{center}
\end{table}

Taking into account systematic errors, our best value for the interaction rate of the 0.862~MeV \ber\ solar neutrinos is 49$\pm$3$_{\rm stat}$$\pm$4$_{\rm syst}$~counts/(day$\cdot$100~ton).  The expected signal for non-oscillated solar $\nu_e$ in the high metallicity SSM~\cite{bib:carlos}\footnote{We remark that in the absence of a resolution between the high-Z abundances reported by Grevesse and Sauval~\cite{bib:gs} and by Asplund, Grevesse, and Sauval~\cite{bib:ags}, for the purpose of comparison with the SSM, we arbitrarily choose as a reference the latest SSM based on the high-Z abundances reported in Ref.~\cite{bib:gs}.  We remark that the current results from Borexino do not help in solving this important controversy.  See Ref.~\cite{bib:bbs} for additional information.} is 74$\pm$4~counts/(day$\cdot$100~ton) corresponding to a flux $\Phi(\ber)$=(5.08$\pm$0.25)$\times$10$^9$~cm$^{-2}$s$^{-1}$.  In the MSW-LMA scenario of solar neutrino oscillation, it is 48$\pm$4~counts/(day$\cdot$100~ton), in very good agreement with our measurement.

Our best estimate for the \kr\ contamination as determined by the fit is 25$\pm$3$_{\rm stat}$$\pm$2$_{\rm syst}$~counts/(day$\cdot$100~ton).  This is consistent with the independent estimate of \kr\ activity from coincidence \kr-\rbm.

Our best estimate for the cosmogenic \cele\ activity induced by cosmic rays at Gran Sasso depth (3800~m.w.e., $\bar{E}_\mu$=320~GeV~\cite{bib:macro}) is 25$\pm$2$_{\rm stat}$$\pm$2$_{\rm syst}$~counts/(day$\cdot$100~ton).  This is 65\% larger than extrapolated from activation on $\mu$ beams at the CERN~NA54 facility~\cite{bib:hagner} and 45\% larger than calculated in Ref.~\cite{bib:11c}.

A minimal extension of the Electroweak Standard Model with a massive neutrino allows a non zero magnetic moment, with the neutrino magnetic moment proportional to the neutrino mass~\cite{bib:marciano}.  The experimental evidence from solar and reactor neutrinos has demonstrated that neutrinos are massive, and may thus possess a non-null magnetic moment.  In the premises of Ref.~\cite{bib:marciano}, the lower limit for the magnetic moment is 4$\times$10$^{-20}$~\mub~\cite{bib:balantekin}.  Larger values are possible in other extensions of the Standard Model~\cite{bib:numu-high}.

In case of a non-null neutrino magnetic moment, the electroweak cross section:
\begin{equation}
\left( \frac{d\sigma}{dT} \right)_W = \frac{2 G^2_F m_e}{\pi} \left[ g^2_L +g^2_R \left( 1 - \frac{T}{E_\nu} \right)^2 - g_L g_R \frac{m_e T}{E_\nu^2} \right]
\end{equation}
is modified by the addition of an electromagnetic term:
\begin{equation}
\left( \frac{d\sigma}{dT} \right)_{EM} = \mu_\nu^2 \frac{\pi \alpha^2_{em}}{m_e^2} \left( \frac{1}{T} - \frac{1}{E_\nu} \right)
\end{equation}
where $E_\nu$ is the neutrino energy and $T$ is the electron kinetic energy.  The shape of the solar neutrino spectrum is sensitive to the possible presence of a non-null magnetic moment, and the sensitivity is enhanced at low energy since $(d\sigma/dT)_{EM}\propto 1/T$.  The SuperKamiokaNDE Collaboration achieved a limit of 1.1$\times$10$^{-10}$~\mub~(90\%~C.L.) using solar neutrino data above a 5-MeV threshold~\cite{bib:beacom,bib:SuperK-munu}.  Ref.~\cite{bib:montanino} presented a limit of 8.4$\times$10$^{-11}$~\mub~(90\%~C.L.) from the \ber\ solar neutrino spectrum in Ref.~\cite{bib:bxfirstresults}.  The best limit on magnetic moment from the study of reactor anti-neutrinos is 5.8$\times$10$^{-11}$~\mub~(90\%~C.L.)~\cite{bib:GEMMA}.

We had previously reported an upper limit of 5.5$\times$10$^{-10}$ using data from the CTF~\cite{bib:bx-munu}.  We now derive bounds on the neutrino magnetic moment by analyzing the $\alpha$-subtracted energy spectrum, obtaining an upper limit of 5.4$\times$10$^{-11}$~\mub~(90\%~C.L.)~\cite{bib:bx-munu2}, which is currently the best experimental limit.

In the MSW-LMA scenario, neutrino oscillations are dominated by matter effects above 3~MeV and by vacuum effects below 0.5~MeV~\cite{bib:mswlma-e}.  The \ber\ neutrinos lie in the lower edge of this transition region.  The measured interaction rate of \ber\ neutrinos depends on the solar $\nu_e$ flux and on the survival probability \Pee\ at the energy of 0.862~MeV.  At present, the only direct measurement of \Pee\ is in the matter-dominated region by observation of $^8$B neutrinos above 5~MeV~\cite{bib:sno}.  The measurement of \Pee\ in and below the transition region is an important test of a fundamental feature of the MSW-LMA scenario.

Under the assumption of the constraint coming from the high metallicity SSM (6\% uncertaintly on \ber\ neutrinos flux), we combine in quadrature systematic and statistical error and we obtain \Pee=0.56$\pm$0.10~(1$\sigma$) at 0.862~MeV.  This is consistent with \Pee=0.541$\pm$0.017, as determined from the global fit to all solar (except Borexino) and reactor data~\cite{bib:kamland}. The no oscillation hypothesis, \Pee=1, is rejected at 4$\sigma$~C.L..
 
Prior to the Borexino measurement the best estimate for \fbe, the ratio between the measured value and the value predicted by the high metallicity SSM~\cite{bib:carlos} for the \ber\ neutrinos flux, was 1.03$^{+0.24}_{-1.03}$~\cite{bib:gonzalez}, as determined through a global fit on all solar (except Borexino) and reactor data, with the assumption of the constraint on solar luminosity.  From our measurement, under the assumption of the constraint from the high metallicity SSM and of the MSW-LMA scenario, we obtain \fbe=1.02$\pm$0.10.  Correspondingly, our best estimate for the flux of \ber\ neutrinos is $\Phi(\ber)$=(5.18$\pm$0.51)$\times$10$^9$~cm$^{-2}$s$^{-1}$.

We then explore the constraint on the flux normalization constants $f_{pp}$ and  \fCNO\ - also defined as the ratio between the measured and predicted values of the respective fluxes - due to the measurement of the \ber\ interaction rate reported in this Letter: the Borexino measured rate can be combined with the other solar neutrino measurements to constrain the flux normalization constants of the other fluxes~\cite{bib:villante}.  The expected rate $R_l$ in the Chlorine and Gallium experiments can be written as:
\begin{equation}
R_l = \sum_i R_{l,i} f_i P_{ee}^{l,i}
\end{equation}
with $l$=\{Ga,Cl\}, $i$=\{$pp$,$pep$,CNO,\ber,\bor\}, $R_{l,i}$ the rate expected in experiment $l$ for source $i$ at the nominal SSM flux, and $P_{ee}^{l,i}$ the survival probability for the source $i$ above the threshold for experiment $l$.  We use $R_{\rm Cl}$=2.56$\pm$0.23~SNU~\cite{bib:rchem-cl}, $R_{\rm Ga}$=68.1$\pm$3.75~SNU~\cite{bib:rchem-ga}, \fB=0.83$\pm$0.07~\cite{bib:sno}, and \fbe=1.02$\pm$0.10, as determined above.

\begin{figure}[!t]
\includegraphics[width=0.5\textwidth]{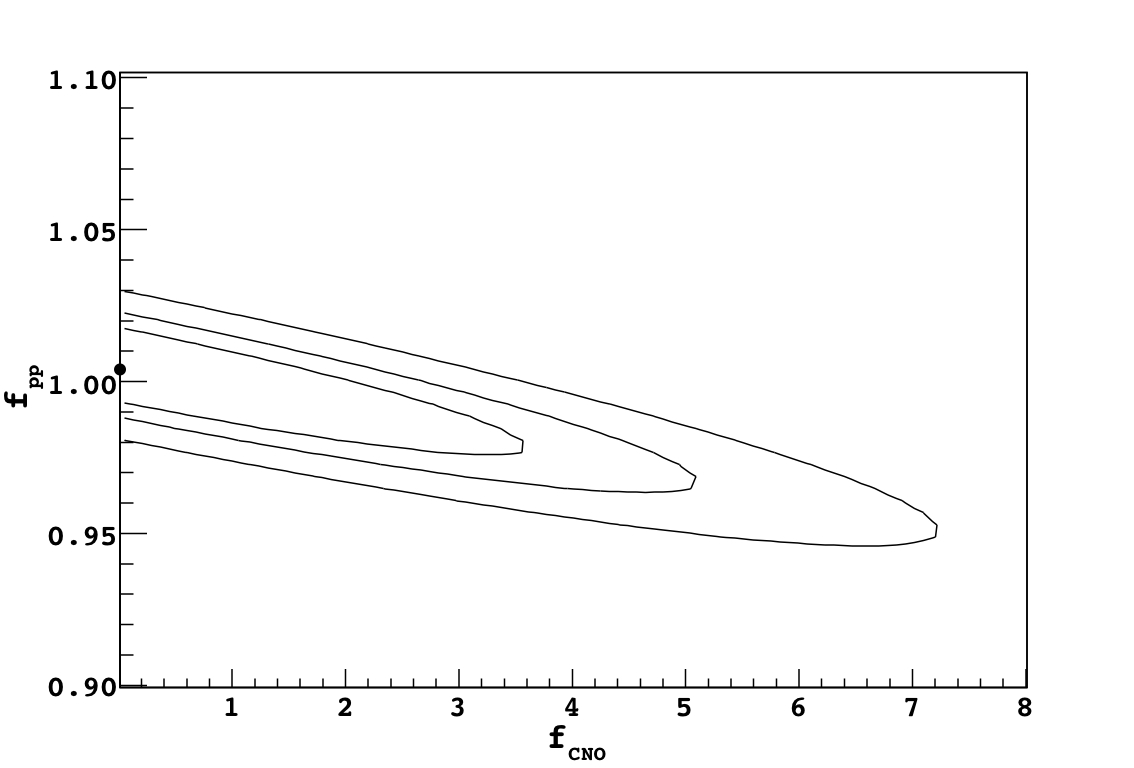}
\caption{Determination of flux normalization constants for {\it pp} and CNO solar neutrinos, \fpp\ and \fCNO\ (68\%, 90\%, and 99\% C.L.).}
\label{fig:fit-pp-CNOs}
\end{figure}

We determine \fpp=1.04$^{+0.13}_{-0.19}$ (1$\sigma$) and \fCNO$<$6.27~(90\%~C.L.) by using the 1-D $\chi^2$-profile method~\cite{bib:pdg}.  The result on \fpp\ represents the best experimental value at present obtained without the luminosity constraint.
The result on \fCNO\ translates into a CNO contribution to the solar luminosity $<$5.4\%~(90\%~C.L.) which is also at present the best limit.  We remark that the SSM we use predicts a CNO contribution on the order of 0.9\%.

Figure~\ref{fig:fit-pp-CNOs} shows the 2-D correlation of \fpp\ and \fCNO\ when adding the luminosity constraint.  Under the same hypothesis, we obtain \fpp=1.005$^{+0.008}_{-0.020}$ (1$\sigma$) and \fCNO$<$3.80 (90\%~C.L.) by using the 1-D $\chi^2$-profile method.  This result on \fpp\ represents the best determination of the $pp$ solar neutrinos flux obtained with the assumption of the luminosity constraint.  The result on \fCNO\ translates into a CNO contribution to the solar neutrino luminosity $<$3.3\%~(90\%~C.L.).

The Borexino program was made possible by funding from INFN (Italy), NSF (USA), BMBF, DFG and MPG (Germany), Rosnauka (Russia), MNiSW (Poland).  We acknowledge the generous support of the Laboratori Nazionali del Gran Sasso (LNGS).  This work has been supported by the ILIAS integrating activity (Contract No.~RII3-CT-2004-506222) as part of the EU FP6 programme.  O.~Smirnov aknowledges the support of Fondazione Cariplo.

\end{document}